\documentclass{elsart}
\usepackage{graphicx}
\usepackage{bm}

\def\lamb#1#2{$^{#1}_{\Lambda}${#2}}
\def\lam#1#2{$^{#1}_{~\Lambda}${#2}}
\def\Kpi{($K^-,\pi^-$) }
\def\Kpig{($K^-,\pi^- \gamma$) }
\def\piK{($\pi^+,K^+$) }

\begin{document}

\begin{frontmatter}

\title{Shell-model calculations for p-shell hypernuclei}

\author{D.J. Millener}
\ead{millener@bnl.gov}
\address{Brookhaven National Laboratory, Upton, NY 11973, USA}

\begin{abstract}
The interpretation of hypernuclear $\gamma$-ray data for p-shell
hypernuclei in terms of shell-model calculations that include
the coupling of $\Lambda$- and $\Sigma$-hypernuclear states
is briefly reviewed. Next, \lamb{8}{Li}/\lamb{8}{Be} and
\lamb{9}{Li} are considered, both to exhibit features of 
$\Lambda$-$\Sigma$ coupling and as possible source of
observed, but unassigned, hypernuclear $\gamma$ rays.
Then, the feasibility of measuring the ground-state doublet
spacing of \lam{10}Be, which, like \lamb{9}{Li}, could be
studied via the $(K^-,\pi^0\gamma)$ reaction, is investigated.
Structural information relevant to the population of states in
these hypernuclei in recent $(e,e'K^+)$ studies is also given.
Finally, the extension of the shell-model calculations to sd-shell
hypernuclei is briefly considered.
\end{abstract}


\end{frontmatter}

\section{Introduction}
\label{intro}

 This article provides an update on the shell-model interpretation 
of $\gamma$-ray transitions in p-shell hypernuclei~\cite{millener08} 
from a previous special issue on recent advances in strangeness 
nuclear physics and the start of an extension to sd-shell
hypernuclei. The experimental data available at the time was 
reviewed by Tamura in the same volume~\cite{tamura08} and
consisted of 22 $\gamma$-ray transitions in \lamb{7}{Li}, \lamb{9}{Be}, 
\lam{11}{B}, \lam{12}{C}, \lam{13}{C}, \lam{15}{N}, and \lam{16}{O},
together with a limit on the ground-state doublet spacing in \lam{10}{B}.
Since then, new results on \lam{11}{B} and \lam{12}{C} from KEK E566 
using an upgraded germanium detector array, Hyperball2, have been 
reported at the Hyp-X conference by Tamura~\cite{tamura10} and 
Ma~\cite{ma10}. The ground-state doublet spacing in \lam{12}{C} is
established as 161 keV both from the direct observation of the 
161 keV $\gamma$-ray and from transitions from an excited $1^-$ state
at 2832 keV. The ground-state doublet spacing is closely related
to that of \lam{10}{B} which is $< 100$ keV. It seems
that the difference between \lam{10}{B} and \lam{12}{C} can
be explained only by invoking the coupling between $\Lambda$ and $\Sigma$
hypernuclear states~\cite{millener10} ($\Lambda$-$\Sigma$ coupling).

  Section~\ref{sec:shell} describes the shell model calculations
and Section~\ref{sec:doublet} summarizes the previously obtained
results for transitions observed in the Hyperball experiments.
Sections~\ref{sec:a8}, \ref{sec:lli9}, and \ref{sec:a10} discuss
the $A\!=\!8$, 9, and 10 hypernuclei. Section~\ref{sec:be} 
contains some information on contributions to the ground-state
binding energies of p-shell hypernuclei. Section~\ref{sec:sd}
is devoted to \lam{19}{F} while Section~\ref{sec:discussion} 
contains a concluding discussion.

\section{Shell-model calculations}
\label{sec:shell}

 Shell-model calculations for p-shell hypernuclei start with the
Hamiltonian 
\begin{equation}
 H = H_N + H_Y + V_{NY} \; ,
\label{eq:hamyn}
\end{equation}
where $H_N$ is an empirical Hamiltonian for the p-shell core,
the single-particle $H_Y$ supplies the $\sim 80$\,MeV mass difference
between $\Lambda$ and $\Sigma$, and $V_{NY}$ is the $YN$ interaction.
The shell-model basis states are chosen to be of the form
$|(p^n\alpha_{c}J_{c}T_{c},j_Yt_Y)JT\rangle$,
where the hyperon is coupled in angular momentum and isospin
to eigenstates of the p-shell Hamiltonian for the core, with up to three
values of $T_c$ contributing for $\Sigma$-hypernuclear states. This
is known as a weak-coupling basis and, indeed, the mixing of
basis states in the hypernuclear eigenstates is generally
very small. In this basis, the core energies are taken from
experiment where possible and from the p-shell calculation otherwise.

 The $\Lambda N$ effective interaction can be  
written~\cite{gsd}
\begin{equation}
V_{\Lambda N}(r)  =  V_0(r) + V_{\sigma}(r)\vec{s}_N\cdot
 \vec{s}_{\Lambda} +  V_{\Lambda }(r)\vec{l}_{N\Lambda }\cdot
\vec{s}_{\Lambda}  + V_{\rm N}(r)\vec{l}_{N \Lambda }\cdot
\vec{s}_{N} +  V_{\rm T}(r)S_{12},
\label{eq:vlam}
\end{equation}
where $S_{12} = 3(\vec{\sigma}_{N}\cdot\vec{r})(\vec{\sigma}_{\Lambda}
\cdot\vec{r})-\vec{\sigma}_{N}\cdot\vec{\sigma}_{\Lambda}$. The
spin-orbit interactions can alternatively be expressed in terms
of the symmetric (SLS) and antisymmetric (ALS) spin-orbit operators
$\vec{l}_{N\Lambda }\cdot (\vec{s}_{\Lambda}\pm\vec{s}_{N})$. The five
$p_N s_\Lambda$ two-body matrix elements depend on the radial integrals 
associated with each component in Eq.~(\ref{eq:vlam}), conventionally denoted 
by the parameters $\overline{V}$, $\Delta$, $S_\Lambda$, $S_N$ and 
$T$~\cite{gsd}. By convention~\cite{gsd}, $S_\Lambda$ and $S_N$ are
actually the coefficients of $\vec{l}_N\cdot\vec{s}_\Lambda$ and
$\vec{l}_N\cdot\vec{s}_N$. Then, the operators associated with
$\Delta$ and $S_\Lambda$ are $\vec{S}_N\cdot \vec{s}_{\Lambda}$
and $\vec{L}_{N}\cdot \vec{s}_{\Lambda}$. 

 The parametrization of Eq.~(\ref{eq:vlam}) applies to the direct 
$\Lambda N$ interaction, the $\Lambda N$--$\Sigma N$ coupling interaction, 
and the direct $\Sigma N$ interaction for both isospin 1/2 and 3/2.
Values for the parameters based on
various Nijmegen models  of the $YN$ interactions are given in Section~3 
of Ref.~\cite{millener10}. Formally, one could include an overall
factor $\sqrt{4/3}\,t_N\cdot t_{\Lambda\Sigma}$ in the analog of 
Eq.~(\ref{eq:vlam}) that defines the interaction, where 
$t_{\Lambda\Sigma}$ is the operator that converts a $\Lambda$ into a 
$\Sigma$. Then, the
core operator associated with  $\overline{V}'$ is $T_N = \sum_i t_{Ni}$.
This leads to a non-zero matrix element only between $\Lambda$ and
$\Sigma$ states that have the same core, with the value
\begin{equation}
 \langle (J_cT,s_\Sigma)JT |V'_{\Lambda\Sigma}|(J_cT,s_\Lambda)JT\rangle
 = \sqrt{4/3}\ \sqrt{T(T+1)}\ \overline{V}'\; ,
\label{eq:fermi}
\end{equation}
in analogy to Fermi $\beta$ decay of the core nucleus. Similarly,
the spin-spin term involves $\sum_i s_{Ni}t_{Ni}$ for the core and 
connects core states that have large Gamow-Teller (GT) matrix elements 
between them. This point has been emphasized by Umeya and Harada~\cite{umeya11}
in a recent article on the effects of $\Lambda$-$\Sigma$ coupling in
$^{7-10}_{\ \ \ \Lambda}$Li. 

 In an LS basis for the core, the matrix elements of 
$\vec{S}_N\cdot \vec{s}_{\Lambda}$ are diagonal
(similarly for $\vec{L}_{N}\cdot \vec{s}_{\Lambda} = 
(\vec{J}_N -\vec{S}_N)\cdot \vec{s}_{\Lambda}$) and
depend just on the intensities of the total $L$ and $S$ 
for the hypernucleus. Because supermultiplet symmetry 
$[f_c]K_cL_cS_cJ_cT_c$ is generally a good symmetry for p-shell core 
states, only one or two values of $L$ and $S$ are important.
The mixing of different $[f_c]L_cS_c$ is primarily due to
the vector (SLS plus ALS) terms in the p-shell Hamiltonian.
Of the remaining $\Lambda N$ parameters, $\overline{V}$ contributes 
only to the overall binding energy; $S_N$ does not contribute to 
doublet splittings in the weak-coupling limit but a negative $S_N$ 
augments the nuclear spin-orbit interaction and contributes to the 
spacings between states based on different core states; in general, 
there are not simple expressions for the coefficients of $T$.

Many hypernuclear calculations have used the venerable
Cohen and Kurath interactions~\cite{ck65}. Here, the p-shell
interaction has been refined using the following strategy.
The one-body spin-orbit splitting between the $p_{3/2}$ and
$p_{1/2}$ orbits is fixed to give a good description
of the light p-shell nuclei (say for $A\leq 9$). The
overall strength of the tensor interaction is also fixed,
ultimately to produce the cancellation in $^{14}$C $\beta$
decay. The well-determined linear combinations of the central 
and vector p-shell interactions are then chosen by fitting
the energies of a large number of states that are known
to be dominantly p-shell in character, including the
large spin-orbit splitting at $A\!=\!15$. A detailed
discussion of p-shell nuclei is given in Section 5 of
Ref.~\cite{millener07}.

\begin{table}
\caption{Doublet spacings in p-shell hypernuclei. Energies are 
given in keV.  The entries in the top 
(bottom) half of the table are calculated using the parameters in 
Eq.~(\ref{eq:param7}) (Eq.~(\ref{eq:param11})). The individual 
contributions do not sum to exactly $\Delta E^{th}$, which comes
from the diagonalization, because small contributions from the
energies of admixed core states are not included.}
\label{tab:spacings}
\begin{tabular*}{\textwidth}{@{}l@{\extracolsep{\fill}}ccrrrrrrr}
\hline\noalign{\smallskip}
 & $J^\pi_u$ & $J^\pi_l$ & $\Lambda\Sigma$ & $\Delta$ & $S_\Lambda$ & $S_N$ 
 & $T$ & $\Delta E^{th}$ & $\Delta E^{exp}$  \\
\noalign{\smallskip}\hline\noalign{\smallskip}
\lamb{7}{Li} & $3/2^+$ & $1/2^+$ & 72 & 628 & $-1$ & 
$-4$ & $-9$ &  693 & 692 \vspace{1pt}\\
\lamb{7}{Li} & $7/2^+$ & $5/2^+$ & 74 & 557 & $-32$ & 
$-8$ & $-71$ &  494 & 471 \vspace{1pt}\\
\lamb{9}{Be} & $3/2^+$ & $5/2^+$ & $-8$ & $-14$ & 
$37$ & $0$ & $28$ &  $44$ & 43 \vspace{5pt}\\
\lam{11}{B} & $7/2^+$ & $5/2^+$ & 56 & 339 & $-37$ & 
$-10$ & $-80$ &  267 & 264 \vspace{1pt}\\
\lam{11}{B} & $3/2^+$ & $1/2^+$ & 61 & 424 & $-3$ & 
$-44$ & $-10$ &  475 & 505 \vspace{1pt}\\
\lam{12}{C} & $2^-$ & $1^-$ & 61 & 175 & $-12$ & 
$-13$ & $-42$ &  153 & 161 \vspace{1pt}\\
\lam{15}{N} & $3/2^+_2$ & $1/2^+_2$ & 65 & 451 & $-2$ & 
$-16$ & $-10$ &  507 & 481 \vspace{1pt}\\
\lam{16}{O} & $1^-$ & $0^-$ & $-33$ & $-123$ & $-20$ & 
1 & 188 &  23 & 26 \vspace{1pt}\\
\lam{16}{O} & $2^-$ & $1^-_2$ & 92 & 207 & $-21$ & 
1 & $-41$ &  248 & 224 \vspace{1pt}\\
\noalign{\smallskip}\hline
\end{tabular*}
\end{table}

\section{Measured doublet spacings}
\label{sec:doublet}

 Table~\ref{tab:spacings} gives a summary of the contributions
from $\Lambda$-$\Sigma$ coupling and the $\Lambda N$ interaction
parameters to all 9 of the measured doublet spacings. Details,
such as figures showing spectra and tables giving breakdowns of 
energy-level spacings, wave functions, and transition rates,
can be found in Refs.~\cite{millener08,millener07}. 
The set of parameters used for \lamb{7}{Li} and \lamb{9}{Be} (chosen
to fit the energy spacings in \lamb{7}{Li}) is (parameters in MeV)
\begin{equation}
\Delta= 0.430\quad S_\Lambda =-0.015\quad {S}_{N}
 = -0.390 \quad {T}=0.030 \; .
\label{eq:param7}
\end{equation}
The  matrix 
elements for the $\Lambda$-$\Sigma$ coupling interaction, based on the 
G-matrix calculations of Ref.~\cite{akaishi00} for the nsc97$e,f$
interactions \cite{rijken99}, are \cite{millener08,millener07}
\begin{equation}
\overline{V}' = 1.45\quad \Delta'= 3.04\quad S_\Lambda' = S_N' = -0.09
\quad T' = 0.16 \; .
\label{eq:paramls}
\end{equation}
These parameters are kept fixed throughout the p-shell in the present 
calculations.

 The ground-state doublet in \lamb{7}{Li} and the exited-state doublets 
in \lam{11}{B} and \lam{15}{N} are based on core states that are 
largely $L_c\!=\!0$ and $S_c\!=\!1$. This limits contributions from other 
than $\Delta$, which enters with a large coefficient of 3/2 in the
$L_c\!=\!0$ limit, and $\Lambda$-$\Sigma$ coupling. The coefficient of
$\Delta$ for the excited state doublet is 7/6 for $L_c\!=\!2$, $S_c\!=\!1$,
$J_c\!=\!3$ but here $S_\Lambda$ and $T$ enter with substantial
coefficients. The $3^+$ core state is the lowest member of an
$L_c\!=\!2$, $S_c\!=\!1$ triplet and moves down in energy if the
strength of the nuclear spin-orbit interaction is increased, as
it is by $S_N$, in the hypernucleus. The value of $S_N$ in 
Eq.~(\ref{eq:param7}) is chosen to reproduce the 2.05-MeV
excitation energy of the $5/2^+$ state in \lamb{7}{Li}.

 The $^8$Be core state for the \lamb{9}{Be} doublet
has mainly $L_c\!=\!2$ and $S_c\!=\!0$, giving a coefficient of $-5/2$
for $S_\Lambda$. The contributions from $\Delta$ and $T$, arising
from small $S_c\!=\!1$ components, together with the small
$\Lambda$-$\Sigma$ coupling contribution, happen to more or less
cancel. This means that the \lamb{9}{Be} doublet spacing demands a 
small value for $S_\Lambda$. 

 The doublet spacings for the heavier p-shell hypernuclei 
consistently require a smaller value for $\Delta$
\begin{equation}
\Delta= 0.330\quad S_\Lambda =-0.015\quad {S}_{N}
 = -0.350 \quad {T}=0.0239 \; .
\label{eq:param11}
\end{equation}
$T$ plays a particularly important role in the ground-state doublet 
($p_{1/2}^{-1}s_\Lambda$) splitting of \lam{16}{O} and is determined 
from a measurement of the doublet spacing \cite{ukai08,ukai04}. 
The ground-state doublet spacing of \lam{15}{N}, which is closely
related to that of \lam{16}{O}, is missing form Table~\ref{tab:spacings}
because the transition to the $1/2^+$ state from the 2268-keV $1/2^+;1$
level is not observed due to a subtle cancellation~\cite{millener08,
millener07}. However, mesonic weak-decay studies have determined
that the ground-state spin-parity of \lam{15}{N} is 
$3/2^+$~\cite{agnello09,gal09}. In Eq.~(\ref{eq:param11}), $S_N$ 
fits the increase in the excitation energy of the excited-state 
doublet over the spacing of the p-hole states in $^{15}$O.

 As can be seen from Table~\ref{tab:spacings}, there is a consistent 
description of the doublet spacings once a larger value of $\Delta$ is 
taken for \lamb{7}{Li}. A conjecture, as yet unproven, is that
shell-model admixtures beyond $0\hbar\omega$ for the lightest
p-shell nuclei ($^6$Li in particular) involve mainly excitations
from the s-shell to the p-shell, thus permitting an active role
for $s_Ns_\Lambda$ matrix elements that are larger than those
for $p_Ns_\Lambda$. For $A\!=\!10$ and beyond, higher admixtures
involve $p\to sd$ excitations and bring in smaller $\Lambda N$
matrix elements.

 Finally, it is clear~\cite{millener08,millener10,millener07} that a
term such as $S_N$ is necessary to describe the spacings between
states based on different core states. Formally, the $S_N$ term arises
from a combination of the SLS and ALS interactions but, in practice,
$S_N$ is treated as a fitting parameter. A two-body $NN$ ALS 
interaction that gives rise to similar effects comes from the
double one-pion exchange $\Lambda NN$ interaction averaged over
the $s_\Lambda$ wave function, as in the original work of Gal, Soper,
and Dalitz~\cite{gsd}. While a one-body $S_N$ term appears to be
adequate near the beginning and end of the p-shell, there is a
need for a much larger effect for (at least) \lam{11}{B}, \lam{12}{C},
and \lam{13}{C}. The high excitation energy (1483 keV) of the
first $1/2^+$ state in \lam{11}{B}, taken together with the
known spacings of the ground-state and first-excited state
doublets, means that the spacing of the doublet centroids is 
1669 keV compared with the $1^+$/$3^+$ core separation of 718 keV.
However, the $S_N$ value of Eq.~(\ref{eq:param11}) gives just over
400 keV towards the difference. While there is sensitivity to
the core wave functions, the high excitation energies of
2832 keV for the $1^-_2$ state of \lam{12}{C} and 4880 keV for the
$3/2^+_1$ state in \lam{13}{C} cannot be explained either
with this value of $S_N$.

\section{The $A\!=\!8$ hypernuclei}
\label{sec:a8}

\begin{figure}[t]
\centering
\includegraphics[width=12.0cm]{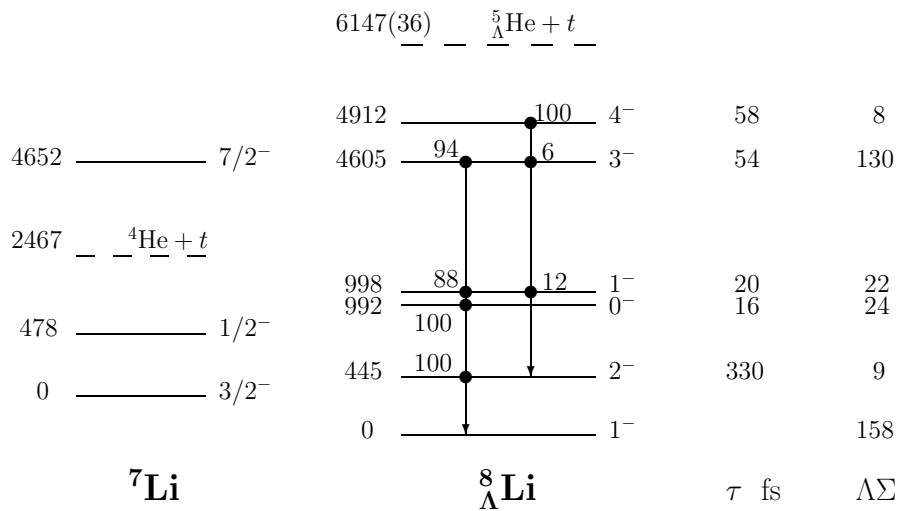}
\caption{The spectrum of \lamb{8}{Li}. The $^7$Li core states 
are shown on the left. The $\gamma$-ray branching ratios and 
lifetimes are theoretical. For each state of \lamb{8}{Li}, the 
calculated energy shifts due to $\Lambda$-$\Sigma$ coupling are 
given. All energies are in keV.}
\label{fig:lli8}
\end{figure}

 Figure~\ref{fig:lli8} gives a theoretical spectrum, including
$\gamma$-ray branching ratios and lifetimes, for \lamb{8}{Li}.
Because the lowest $3/2^-$ and $1/2^-$ states of $^7$Li are
closely spaced and both have $L_c\!=\!1$, the bound $1^-$ states of 
\lamb{8}{Li} involve significant mixing  ($|1^-_1\rangle = 
0.946\,(3/2^-)\times s_\Lambda -0.319\,(1/2^-)\times s_\Lambda$) 
of the configurations based on them. On a historical note,
the spin-parity of \lamb{8}{Li}~\cite{davis63} and restrictions 
on the mixing~\cite{bohm74}  were derived from emulsion data on 
the decay \lamb{8}{Li}$\to \pi^-$ + $^8{\rm Be}^*\to \pi^- 
+\alpha + \alpha$. The next largest admixtures are actually from 
the corresponding $\Sigma$-hypernuclear states.

 As can be see from Fig.~\ref{fig:lli8}, the energy shift
due $\Lambda$-$\Sigma$ coupling for the $1^-$ ground state
is large in strong contrast to that for the $2^-$ member of
the doublet. In fact, $\Lambda$-$\Sigma$ coupling is predicted
to account for a third of the doublet spacing. The predicted
energy spacing is very close to the 442.1(21) keV energy of a 
$\gamma$-ray observed following the production of highly-excited states
of \lam{10}{B} via the $(K^-,\pi^-)$ reaction~\cite{chrien90}.
The 442 keV $\gamma$-ray was tentatively attributed to \lamb{8}{Li}
or its mirror hypernucleus \lamb{8}{Be}~\cite{chrien90}, the $2^-$ 
state of which could be reached by $l\!=\!1$ deuteron emission
from the $s^4p^4(sd)s_\Lambda$ component of the 
$3^+$ $s_n^{-1}s_\Lambda$ substitutional state in 
\lam{10}{B} (the excited-state doublet in \lamb{7}{Li} was
studied following $l\!=\!0$ $^3$He emission from the same
state~\cite{ukai06}). If confirmed, the 442-keV $\gamma$-ray 
would provide additional strong support for the important role 
played by $\Lambda$-$\Sigma$ coupling in hypernuclear spectra.

 A 1.22(4)-MeV $\gamma$-ray, seen after $K^-$ mesons were stopped 
in a $^9$Be target~\cite{bejidian80}, was tentatively ascribed to \lamb{8}{Li},
possibly as a transition between the $1^-$ states in Fig.~\ref{fig:lli8}.
In Ref.~\cite{mgdd85}, it was found to be difficult to explain
such a high energy. This is still the case despite the addition of
136 keV to the transition energy from $\Lambda$-$\Sigma$ coupling.

 The role of $\Lambda$-$\Sigma$ coupling for the ground-state
doublet and the excited $1^-$ state can be seen from the interplay
of $\overline{V}'$ and $\Delta'$ in the coupling matrix elements
and the use of perturbation theory for the energy shifts ($\sim
v^2/{\Delta E}$ with $\Delta E\sim 80$ MeV). From Eq.~(\ref{eq:paramls}),
$\overline{V}'$ gives a contribution of 1.45 MeV to the diagonal coupling
matrix elements involving the same core state. Adding the remaining
$\Lambda$-$\Sigma$ contributions (mainly from  $\Delta'$) for
the $3/2^-$ core states gives 2.6524 MeV for $J^\pi\!=\!1^-$ and
0.7286 MeV for $J^\pi\!=\!2^-$. The off-diagonal matrix elements
(from $\Delta'$) involving $1/2^-$ and $3/2^-$ core states are
both $-1.5094$ MeV, while the diagonal $1/2^-$ matrix element is
1.4761 MeV. In the weak-coupling limit, the push on the ground
state is 88 keV from the $(3/2^-\times s_\Sigma)$ state and 28 keV 
from the $(1/2^-\times s_\Sigma)$ state while the push on the $2^-$
state is only 7 keV. The mixing between the $(3/2^-\times s_\Lambda)$
and $(1/2^-\times s_\Lambda)$ states increases the $\Lambda$-$\Sigma$ 
coupling matrix elements for the lower $1^-$ state and decreases
them for the upper state; putting in the numbers from above, an
estimate of 157 keV is obtained for the push on the ground state.

 As far as electromagnetic transitions are concerned, the
p-shell wave functions account well for the M1 properties of the
$3/2^-$ and $1/2^-$ core states using the bare M1 operator, leaving 
room for the expected small enhancement of the isovector matrix 
elements by meson-exchange currents~\cite{marcucci08}; the calculated 
magnetic moments of $^7$Li and $^7$Be are 3.145 $\mu_N$ and
$-1.263$ $\mu_N$ compared with the experimental values of 3.256 $\mu_N$
and $-1.399$ $\mu_N$, respectively. Because the M1 matrix element for the
ground-state doublet transition is proportional to $g_c -g_\Lambda$
 ($g_\Lambda\!=\!-1.226$) in the weak-coupling limit~\cite{dg78}
the transition is going to be much faster in the odd-proton 
nucleus \lamb{8}{Li}. This simple approximation does not apply
because of the configuration mixing in the $1^-$ wave functions.
The M1 transition strengths end up being 1.086 W.u. in \lamb{8}{Li}
and 0.043 W.u. in \lamb{8}{Be} ($\tau\!=\!8.37$ ps). For the 
decay of the $1^-_2$ state in \lamb{8}{Be}, one obtains  a
branch of 64\% to the ground state and a lifetime of 46 fs.
All these low-energy transitions are little affected by
their E2 components.

 As a further example of the effects of $\Lambda$-$\Sigma$ coupling,
the \lamb{8}{He} ground-state doublet is predicted to have a
spacing of 101 keV with $\Lambda$-$\Sigma$ contributions of 154
and 182 keV to the binding energies of the $1^-$ and $2^-$ states,
respectively. In this case, $\Lambda$-$\Sigma$ coupling  reduces 
the doublet spacing by 28 keV.

\section{The \lamb{9}{Li} hypernucleus}
\label{sec:lli9}

 There is interest in \lamb{9}{Li} because it has been studied
using the $^9$Be$(e,e'K^+)$\lamb{9}{Li} reaction
at JLab~\cite{cusanno10} and could be studied via the  
$^9$Be$(K^-,\pi^0\gamma)$\lamb{9}{Li} reaction at J-PARC.
In addition, it is the possible source of a 1303-keV $\gamma$-ray
seen  in a stopped $K^-$ experiment~\cite{miwa05}, 
most strongly on a $^9$Be target.

\begin{figure}[t]
\centering
\includegraphics[width=12.0cm]{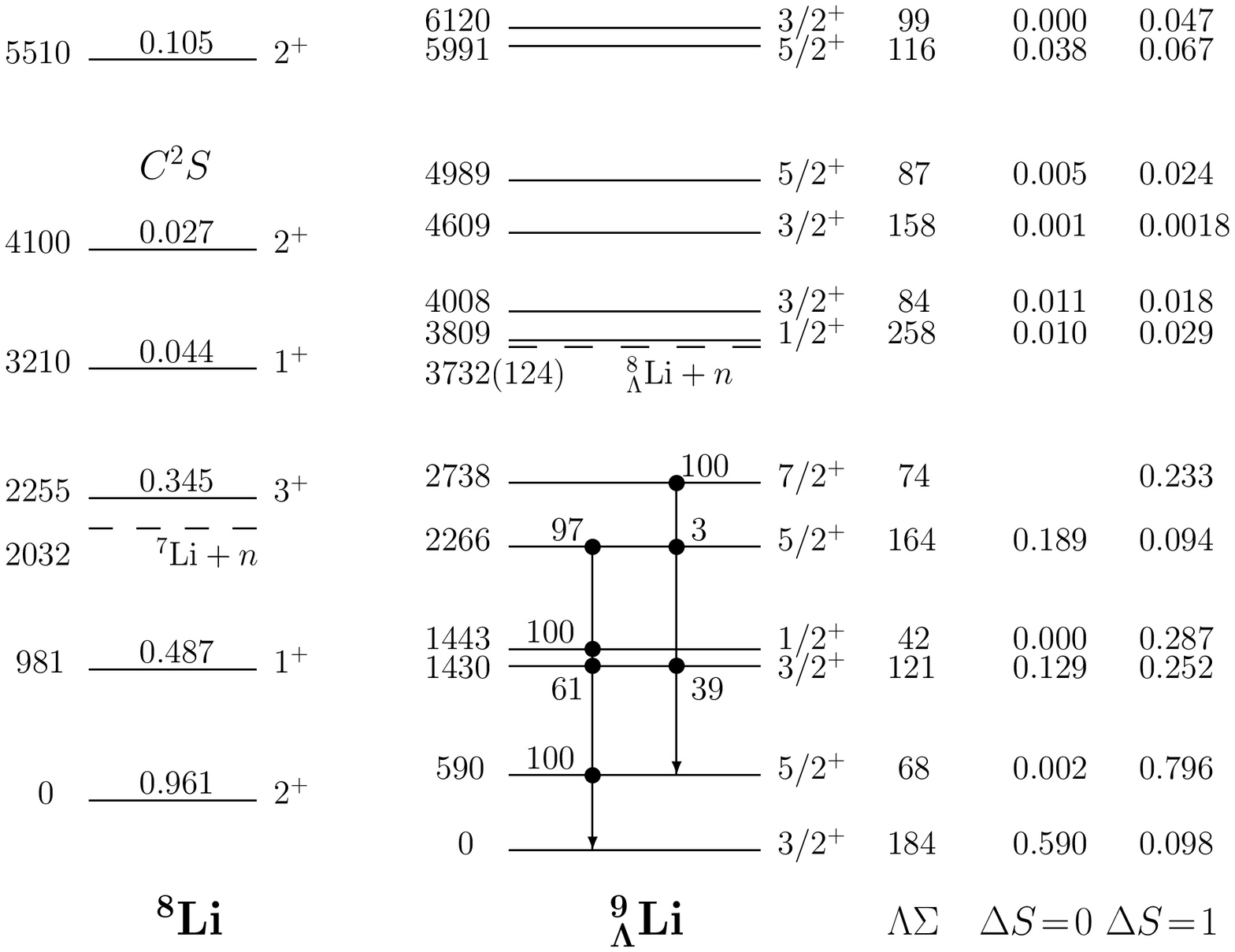}
\caption{The spectrum of \lamb{9}{Li}. The $^8$Li core states 
are shown on the left along with the spectroscopic factors for
proton removal from $^9$Be. The $\gamma$-ray branching ratios and 
lifetimes are theoretical. For each state of \lamb{9}{Li}, the 
calculated energy shifts due to $\Lambda$-$\Sigma$ coupling are 
given. All energies are in keV. On the right, the structural
factors (defined in Appendix A) giving the 
relative population of levels in purely non-spin-flip 
($\Delta S\!=\!0$) and purely spin-flip ($\Delta S\!=\!1$) 
production reactions on a $^9$Be target are given.}
\label{fig:lli9}
\end{figure}

 Figure~\ref{fig:lli9} gives a theoretical spectrum, including
$\gamma$-ray branching ratios and lifetimes, for \lamb{9}{Li}.
Because non-spin-flip production is dominant in the $(K^-,\pi^0)$
reaction at rest, the only likely candidate for the 1303-keV
$\gamma$-ray is the excited $3/2^+$ to ground state transition. In 
this case, an 840-keV transition to the lowest $5/2^+$ state should
also be observable. The predicted energy of 1430 keV is too high.
For comparison, the $\Lambda N$ parameter set in Eq.~(\ref{eq:param11})
predicts the $3/2^+_2$ state at 1331 keV and the $5/2^+_1$ state
at 471 keV. It is clear that an in-flight $(K^-,\pi^0\gamma)$
study with the Hyperball-J at an incident $K^-$ energy where 
spin-flip amplitudes are important is desirable.

 From the $^9$Be$(t,\alpha)^8$Li study by Liu and Fortune~\cite{liu88}
and the pickup spectroscopic factors given in Fig.~\ref{fig:lli9},
the bulk of the cross section for the production of $s_\Lambda$
states in the  $^9$Be$(e,e'K^+)$\lamb{9}{Li} reaction is expected
to be concentrated in states built on the lowest three states of
$^8$Li (see Ref.~\cite{sotona94} for an early theoretical study). 
This is indeed the case~\cite{cusanno10}. The strongest
observed state is the upper member of the ground-state doublet
but there appears to be more strength in the states based on the
$3^+$ core state than predicted and a disagreement about the location
of the strength based on the $1^+$ state. The strength based on the 
$1^+_2$ state is close enough to the neutron threshold that the
states should be narrow and any significant strength associated
with them should be observable in the electro-production reaction. 
This would be the case for the (8-16)2BME and (8-16)POT interactions
of Cohen and Kurath~\cite{ck65} but not for the (6-16)2BME
interactions or the various fitted interactions used in recent
hypernuclear studies. The former interactions favor the second
$1^+$ state in proton removal from $^9$Be because the lowest
$1^+$ state is dominantly $L_c\!=\!1$, $S_c\!=\!1$ rather
than strongly mixed $S_c\!=\!0$ and $S_c\!=\!1$. The use of
the (8-16)2BME interaction  is the reason
that the \lamb{9}{Li} spectrum of Umeya and Harada~\cite{umeya11}
looks rather different from the one in Fig.~\ref{fig:lli9}.

\section{The $A\!=\!10$ hypernuclei}
\label{sec:a10}

\begin{figure}[t]
\centering
\includegraphics[width=12.0cm]{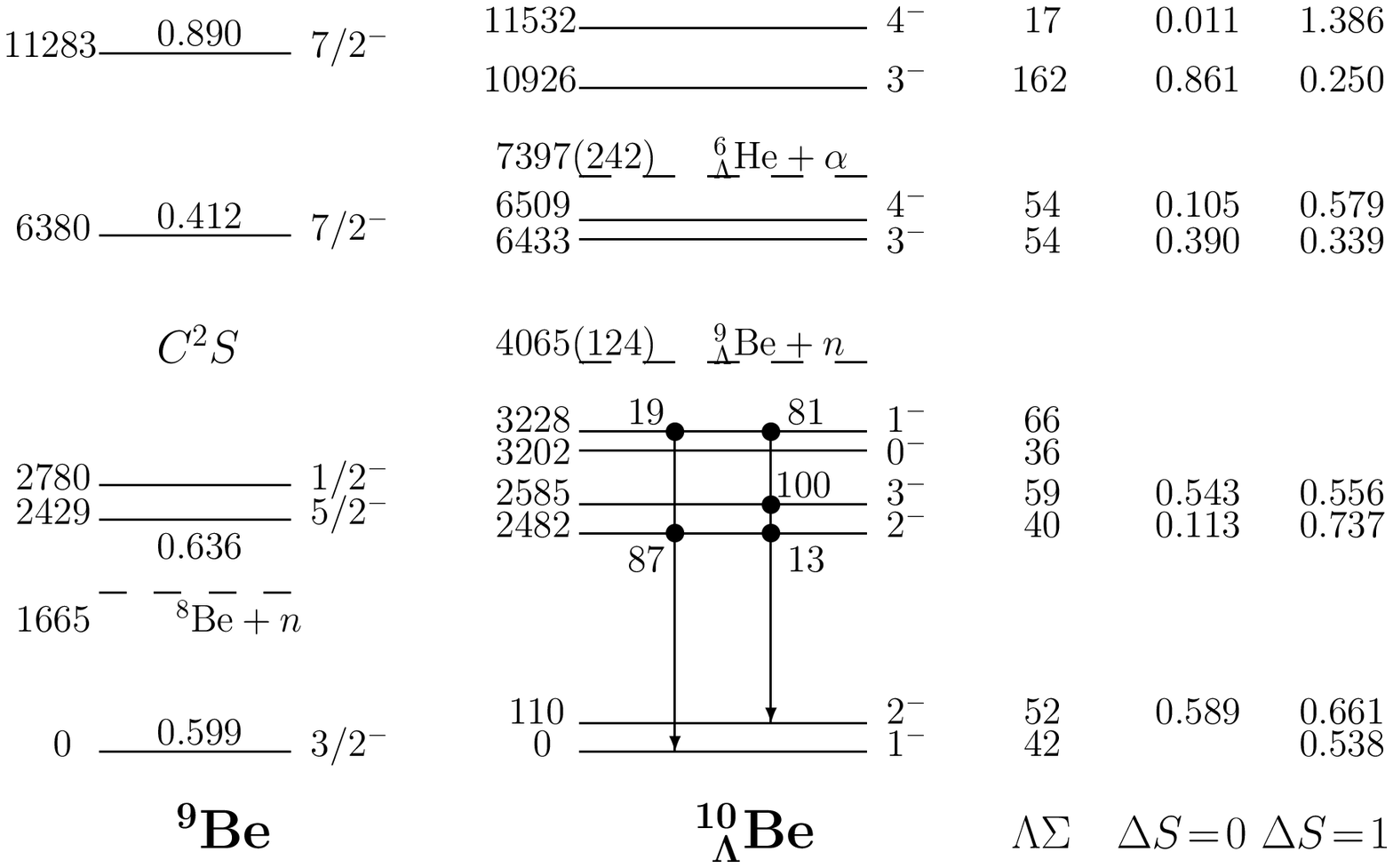}
\caption{The spectrum of \lam{10}{Be}. The $^9$Be core states 
are shown on the left along with the spectroscopic factors for
proton removal from $^{10}$B. The $\gamma$-ray branching ratios and 
lifetimes are theoretical. For each state of \lam{10}{Be}, the 
calculated energy shifts due to $\Lambda$-$\Sigma$ coupling are 
given. All energies are in keV. On the right, the structural
factors giving the relative population of levels in purely
non-spin-flip ($\Delta S\!=\!0$) and purely spin-flip 
($\Delta S\!=\!1$) production reactions on a $^{10}$B target
are given.}
\label{fig:lbe10}
\end{figure}

 Figure~\ref{fig:lbe10} gives a theoretical spectrum, including
$\gamma$-ray branching ratios, pickup spectroscopic factors,
and formation factors for \lam{10}{Be}.

 \lam{10}{Be} is another hypernucleus that could be studied via the  
$(K^-,\pi^0\gamma)$ reaction with the Hyperball-J at J-PARC, this 
time with a $^{10}$B target. A strong reason for doing so would be 
to try to measure the ground-state doublet spacing by observing
transitions to both members from a higher level. The obvious candidate
is the $2^-$ level based on the $5/2^-$ core level of $^9$Be.
Unfortunately, the $2^-\to 2^-$ transition is strongly hindered with
respect to the $2^-\to 1^-$ transition by a factor of 15 from the 
recoupling coefficient (but gains something back on the 
$2J_f\!+\!1$ factor). In the weak-coupling limit, the 
$2^-\to 2^-$ branch would be only 9\% but something is gained 
from configuration mixing. Again, the E2 components of the 
transitions are not very important.
 
 The spacings of the ground-state and excited-state doublets 
are predicted to be very similar. This could certainly be checked
in a $(K^-,\pi^0\gamma)$ experiment and which $\gamma$-ray is
which could be determined by choosing $K^-$ momenta for
which the ratio of spin-flip to non-spin-flip is quite different.
The $1^-$ level based on the broad $1/2^-$ level in $^9$Be
could be populated via the \Kpi\ reaction on $^{10}$Be if only
a thick enough $^{10}$Be target could be made.

  As far as production reactions are concerned, the 
$^{10}$B\Kpi\lam{10}{B} reaction has been studied in 
KEK E336~\cite{hashtam06}. From the spectroscopic factors for
proton or neutron removal from $^{10}$B, one expects to
see four strong peaks up to about 10 MeV in excitation energy.
In the \Kpi\ experiment, these are not cleanly resolved but
the data has been fitted to extract energies and relative 
yields~\cite{hashtam06}. The relative yields are in good agreement
with the $\Delta S\!=\!0$ structure factors in Fig.~\ref{fig:lbe10}
but the spectrum is somewhat expanded with respect to that
extracted from the data (the extracted energy may be affected
by the steeply rising background that extends under the fourth
peak). A spectrum in which the four peaks, based on the core states
reached strongly by proton removal from $^{10}$B,  are cleanly separated
has recently been obtained using the $^{10}$B$(e,e'K^+)$\lam{10}{Be} 
reaction at JLab~\cite{tang12}. This is as predicted by Motoba et 
al.~\cite{motoba94} and by the results in Fig.~\ref{fig:lbe10} based
on more recent information on in-medium $YN$ interactions. States 
involving a $p_\Lambda$ coupled to the same core states are also 
expected to be strongly populated (again see Ref.~\cite{motoba94}) 
and it will be interesting to make a detailed comparison between 
theory and experiment.

\section{Contributions to $\Lambda$ binding energies}
\label{sec:be}

\begin{table}[t]
\caption{$\Lambda$-$\Sigma$ and spin-dependent contributions to 
ground-state binding energies (in keV). The units for $B_\Lambda$ and 
$\overline{V}$ in the last two columns are MeV. The experimental
$B_\Lambda$ values and errors are taken from Ref.~\cite{davis86}. 
No $\overline{V}$ is given for \lamb{9}{Be} which is a special case 
because of the unbound nature of $^8$Be, the binding energy
of which enters into $B_\Lambda$(\lamb{9}{Be}).
The entry for \lam{12}{Be} is based on p-shell core states but the 
ground state could have positive parity.}
\label{tab:be}
\begin{tabular*}{\textwidth}{@{}c@{\extracolsep{\fill}}rrrrrrrrr}
\hline\noalign{\smallskip}
 & $J^\pi$ & $\Lambda$-$\Sigma$ & $\Delta$ & $S_\Lambda$ & $S_N$ &
 $T$ & Sum & $B_\Lambda^{expt}(\Delta B)$ & $\!\!\overline{V}$ \\
\hline\noalign{\smallskip}
\lamb{7}{He} & $1/2^+$ & 101 & 1 & 0 & 176 & 0 & 278 &  &  \\
\lamb{8}{He} & $1^-$ & 154 & 152 & $-13$ & 454 & $-38$ & 709 & 7.16(70) & $-1.11$ \\
\lamb{9}{He} & $1/2^+$ & 253 & 6 & 0 & 619 & 1 & 879 &  &  \\
\lamb{7}{Li} & $1/2^+$ & 78 & 419 & 0 & 94 & $-2$ & 589 & 5.58(3) & $-0.94$ \\
\lamb{8}{Li} & $1^-$ & 160 & 288 & $-6$ & 192 & $-9$ & 625 & 6.80(3) & $-1.02$ \\
\lamb{9}{Li} & $3/2^+$ & 183 & 350 & $-10$ & 434 & $-6$ & 952 & 8.50(12) & $-1.06$ \\
\lam{10}{Li} & $1^-$ & 275 & 175 & $-11$ & 595 & $-12$ & 1022 &  &  \\
\lamb{9}{Be} & $1/2^+$ & 4 & 0 & 0 & 207 & 0 & 211 & 6.71(4) & \\
\lam{11}{Be} & $1/2^+$ & 99 & 2 & 0 & 540 & 0 & 641 &  &  \\
\lam{12}{Be} & $0^-$ & 158 & $-76$ & $-15$ & 554 & 127 & 748 &  &  \\
\lam{10}{B} & $1^-$ & 35 & 125 & $-13$ & 386 & $-15$ & 518 & 8.89(12) & $-1.05$ \\
\lam{11}{B} & $5/2^+$ & 66 & 203 & $-20$ & 652 & $-43$ & 858 & 10.24(5) & $-1.04$ \\
\lam{12}{B} & $1^-$ & 103 & 108 & $-14$ & 704 & $-29$ & 869 & 11.37(6) & $-1.05$ \\
\lam{13}{B} & $1/2^+$ & 130 & 197 & $-6$ & 621 & $-57$ & 885 &  &  \\
\lam{14}{B} & $1^-$ & 255 & 115 & $-13$ & 458 & $-30$ & 785 &  &  \\
\lam{13}{C} & $1/2^+$ & 28 & $-4$ & 0 & 841 & $-1$ & 864 & 11.69(12) & $-0.96$ \\
\lam{14}{C} & $1^-$ & 75 & 47 & 6 & 816 & $-40$ & 904 & 12.17(33) & $-0.91$ \\
\lam{15}{C} & $1/2^+$ & 116 & 8 & 0 & 636 & 2 & 762 &  &  \\
\lam{15}{N} & $3/2^+$ & 59 & 40 & 12 & 630 & $-69$ & 726 & 13.59(15) & $-0.97$ \\
\lam{16}{N} & $1^-$ & 62 & 94 & 6 & 349 & $-45$ & 412 & 13.76(16) & $-0.93$ \\
\noalign{\smallskip}\hline
\end{tabular*}
\end{table}

 Table~\ref{tab:be} shows the $\Lambda$-$\Sigma$ and spin-dependent 
contributions to the ground-state binding energies for a wide
range of p-shell hypernuclei. The sum of these contributions
can reach 1 MeV. The experimental $B_\Lambda$ values are from
emulsion studies~\cite{davis86} (a number of mirror hypernuclei
are not listed) except for \lam{16}{N} where the value comes 
from a study using the $^{16}$O$(e,e'K^+)$\lam{16}{N}
reaction~\cite{cusanno09} (the observed $1^-$ state is actually
26 keV above the $0^-$ ground state~\cite{ukai08,ukai04}).
The remaining hypernuclei that are listed are chosen because
the neutron-rich p-shell cores have higher isospin and can exhibit 
larger effects from $\Lambda$-$\Sigma$ coupling.

 The spin-independent central component $\overline{V}$ of the
$\Lambda N$ interaction doesn't affect the spectra but can be
estimated from the binding energies by taking 
$B_\Lambda$(\lamb{5}{He}) = 3.12 MeV as the $s_\Lambda$ 
single-particle energy and using
\begin{equation}
 B_\Lambda = 3.12 - n\, \overline{V} + {\rm Sum}\, ,
\label{eq:vbar}
\end{equation}
where $n$ is the number of p-shell nucleons in the core.
The values of $\overline{V}$ so extracted are given in the
last column of Table~\ref{tab:be}.
The $\overline{V}$ are relatively constant and close to the values
derived from $\Lambda N$ potential models~\cite{millener10}.
In reality, small repulsive contributions quadratic in $n$
are expected from the double one-pion exchange interaction~\cite{gsd}
which would then call for a somewhat more attractive $\overline{V}$.
Quite good estimates can be made for $B_\Lambda$ values. In
the case of $\Lambda\Lambda$ hypernuclei, two spin-averaged 
$B_\Lambda$ values enter into the binding energy along with
a $\Lambda\Lambda$ two-body matrix element that is known to
be quite small ($-0.67$ MeV). Then, it is clear that the
knowledge of single-$\Lambda$ binding energies can be used
to make reliable estimates for the binding energies of
$\Lambda\Lambda$ hypernuclei~\cite{gal11}.

\section{sd-shell hypernuclei}
\label{sec:sd}

 An extension of the studies of $\gamma$-ray transitions in
p-shell hypernuclei is planned for \lam{19}{F} in J-PARC 
E13~\cite{tamura12}. The reason for choosing $^{19}$F as a
target is that $^{18}$F has a primarily $L\!=\!0$, $S\!=\!1$ 
ground state so that one should observe a relatively large
ground-state doublet spacing for \lam{19}{F}, in complete
analogy to \lamb{7}{Li} (and the \lam{15}{N} doublet based on the
second $1^+$ state in $^{15}$N); $^{19}$F itself has a primarily
$L\!=\!0$, $S\!=\!1/2$ ground state, which is why an (impractical)
$^{20}$Ne target was considered by Millener et al.~\cite{mgdd85}.

 The $^{18}$F core nucleus has quite a dense spectrum, including
$3^+;0$, $0^+;1$, $0^-;0$, and $5^+;0$ states close to 1 MeV
(see Fig.3 in Ref.~\cite{tamura12}). The wave functions for the
lowest $1^+$ and $3^+$ states are given in a $jj$-coupling basis
in Table~\ref{tab:cw}. The $1^+$ state is actually 92.7\%
$L\!=\!0$, $S\!=\!1$ (amplitudes 0.8985 and $-0.3461$ for SU(3) 
symmetry $(4\,0)$ and $(0\,2)$, respectively), while the $3^+$
state is 96.9\%  $L\!=\!2$, $S\!=\!1$ (amplitudes 0.9722 and 
$-0.1548$). Historically, this simplicity for $^{18}$F and 
$^{19}$F was a significant factor in the introduction of Elliott's
SU(3) model~\cite{elliott99}.

\begin{table}[t]
\caption{ $^{18}$F wave functions using the Chung-Wildenthal
interaction~\cite{chung76}.}
\label{tab:cw}
\begin{tabular*}{\textwidth}{@{}c@{\extracolsep{\fill}}rrrrrr}
\hline\noalign{\smallskip}
 $J^\pi$ & $d_{5/2}^2$ & $d_{5/2}d_{3/2}$ & $d_{3/2}^2$ &
$d_{5/2}s_{1/2}$ & $d_{3/2}s_{1/2}$ & $s_{1/2}^2$ \\
\hline\noalign{\smallskip}
  $1^+$ & 0.6038 & $-0.6539$ & $-0.0515$ & & 0.1130 & 0.4386 \\
  $3^+$ & 0.5721 & $-0.2397$ & $-0.0026$ & 0.7844 &  &  \\
\noalign{\smallskip}\hline
\end{tabular*}
\end{table}

\begin{table}
\caption{ $(sd)_Ns_\Lambda$ matrix elements for harmonic
oscillator (HO) and Woods-Saxon (WS) wave functions. The $s_\Lambda$
orbit is bound at 15 MeV. The sd orbits are bound as indicated,
where Exp. means binding energies of 8, 4, and 9 MeV for the
$1s_{1/2}$, $0d_{3/2}$, and $0d_{5/2}$ neutron orbits, taking into
account that the neutron separation energy is 9.15 MeV for $^{18}$F.
The well geometry has $r_n\!=\!1.26$ fm and $a_n\!=\!0.65$ fm
for the neutron (1.212 fm and 0.60 fm for the $\Lambda$).}
\label{tab:sd}
\begin{tabular*}{\textwidth}{@{}c@{\extracolsep{\fill}}rrrrr}
\hline\noalign{\smallskip}
 &  & HO  & WS  & WS & WS \\
 & $J$ & $b\!=\!1.741$ fm & ${\rm BE}\!=\!{\rm Exp.}$ & 
${\rm BE}\!=\!9$ MeV & ${\rm BE}\!=\!1$ MeV \\
\hline\noalign{\smallskip}
 $\langle 1s_{1/2}s_\Lambda|V|1s_{1/2}s_\Lambda\rangle$ & 0 &
$-1.6067$ & $-1.2774$ & $-1.3181$ & $-0.6529$ \\
 & 1 & $-1.1817$ & $-0.9524$ & $-0.9822$ & $-0.4915$ \\
 $\langle 0d_{3/2}s_\Lambda|V|1s_{1/2}s_\Lambda\rangle$ & 1 &
$-0.1254$ & $-0.1062$ & $-0.1174$ & $-0.0610$ \\
 $\langle 0d_{3/2}s_\Lambda|V|0d_{3/2}s_\Lambda\rangle$ & 1 &
$-0.4883$ & $-0.4890$ & $-0.5522$ & $-0.3828$   \\
 & 2 & $-0.5184$ & $-0.5107$ & $-0.5747$ & $-0.4033$ \\
 $\langle 0d_{5/2}s_\Lambda|V|0d_{3/2}s_\Lambda\rangle$ & 2 &
$0.1301$ & $0.1333$ & $0.1459$ & $0.0936$ \\
 $\langle 0d_{5/2}s_\Lambda|V|0d_{5/2}s_\Lambda\rangle$ & 2 &
$-1.0508$ & $-1.0708$ & $-1.1026$ & $-0.7453$ \\
 & 3 & $-0.9863$ & $-1.0009$ & $-1.0309$ & $-0.6958$ \\
\noalign{\smallskip}\hline
\end{tabular*}
\end{table}

 In the sd-shell, there are 8 $(sd)_Ns_\Lambda$ matrix elements;
4 central, one each for LS and ALS in relative p states, and 2
tensor (in both even and odd states). These are shown as a function 
of binding energy in Table~\ref{tab:sd}. Here, a radial representation
of the $\Lambda N$ interaction that reproduces the matrix elements
of Eq.(\ref{eq:param11}) is used (cf. Ref.~\cite{millener10}).
Table~\ref{tab:sd} demonstrates that the matrix elements are
sensitive to the binding energies of the sd-shell orbits,
especially the noded $1s$ orbit. We note that sd-shell orbits 
indeed become loosely bound and the $1s$ orbit moves below the
$0d_{5/2}$ orbit for states in p-shell hypernuclei.

  Combining the Woods-Saxon matrix elements for the Exp. case
in  Table~\ref{tab:sd} with the wave functions in Table~\ref{tab:cw},
the doublet spacings for states based on the lowest $1^+$ and $3^+$
states are 305 keV and 196 keV, respectively. This calculation is
for simple weak-coupling states without the inclusion of
$\Lambda$-$\Sigma$ coupling and calculations similar
to those performed for p-shell hypernuclei remain to be 
performed. In addition, $^{18}$F has low-lying negative parity
states that can be reached in the \Kpig\ reaction~\cite{tamura12}. 
The lowest $0^-$ and $1^-$ states are predominantly of the form 
$p_{1/2}^{-1}\times {^{19}{\rm F}(gs)}$ (89\% and 81\% for
$0^-$ and $1^-$ in a full $1\hbar\omega$ shell-model calculation;
alternatively 72.4\% and 74.6\% $(6\,1)$ SU(3) symmetry). The 
$1\hbar\omega$ hypernuclear basis requires  $(sd)^2p_\Lambda$ 
states in addition to $p^{-1}(sd)^3s_\Lambda$ and $(sd)(pf)s_\Lambda$
to make a non-spurious basis.

\section{Discussion}
\label{sec:discussion}

 Calculations using refined ($0\hbar\omega$) interactions for
p-shell core nuclei, with the inclusion of $\Lambda$-$\Sigma$
coupling, have been quite successful in that a large body of
data on hypernuclear level spacings has been correlated with
relatively few $YN$ parameters. The introduction of explicit
$\Lambda$-$\Sigma$ coupling is generally beneficial. It is
necessary to understand the s-shell hypernuclear binding
energies, especially for \lamb{4}{H} and \lamb{4}{He}.
It makes significant contributions of varying size
relative to the dominant $\Lambda N$ spin-spin interaction
in p-shell doublet spacings and binding energies. In particular,
it seems necessary to understand the ground-state doublet spacings
in both \lam{10}{B} (a limit) and \lam{12}{C}. These effects involve
the interplay of Fermi and Gamow-Teller type matrix elements
connecting core states.

 The remaining problems are to understand (1) the need for
different $\Lambda N$ spin-spin interaction strengths
at either end of the p-shell and (2) the need for a stronger
enhancement of the nuclear vector interaction terms (LS and
ALS) near mid shell by the presence of the $\Lambda$.
The next steps are to expand the shell-model basis and to
reintroduce the double one-pion exchange $\Lambda NN$ 
interaction considered by Gal, Soper, and Dalitz~\cite{gsd}.

For consistency, one has to go beyond $2\hbar\omega$ states
for the core and the $\Lambda$ configurations which makes
for a challenging problem. A somewhat more tractable
problem is to treat the full $1\hbar\omega$ basis of 
hypernuclear states~\cite{zofka91}. This is necessary to (1) treat
properly $p_\Lambda$ states in both the p- and sd-shells and
(2) estimate decay widths for particle emission from
unbound hypernuclear states~\cite{majling92,majling97}, 
this being the way in which $\gamma$ transitions in daughter 
hypernuclei have been studied.

\section*{Acknowledgements}

This work has been supported by the US Department of Energy under Contract
No. DE-AC02-98CH10886 with Brookhaven National Laboratory.

\section*{Appendix A. Structure factors for production reactions}
\label{sec:appendix}

 For a particular $l_n\to l_\Lambda$ transition, it is possible
to pull out a structure factor that multiplies a particular 
distorted (or plane) wave radial integral and governs the
relative cross sections for related states. See, for example, Section
3.2 of Ref.~\cite{auerbach83} where the structure factor is 
$(2J_f+1)/(2J_i+1)$ times the square of an LS one-body density-matrix
(OBDME) for the transition, together with some common factors
such as the square of the isospin Clebsch-Gordan coefficient for
the transition. For the predominantly non-spin-flip transitions
in \Kpi\ or \piK\ reactions, there is a single OBDME with $S\!=\!0$.
In the case $(e,e'K^+)$ reactions, the Kroll-Ruderman term
${\bf\sigma}\cdot {\bf\epsilon}$ is dominant and one needs to evaluate
the magnetization current contributions to the transverse electric and
magnetic operators that appear in the $(e,e')$ cross section,
specifically the $\Sigma$ and $\Sigma'$ terms that appear in
Eqs.(22b) and (22c), and given in Eqs.(1d) and (1e), of
Donnelly and Haxton~\cite{donnelly79}. For the same L, we
can pull out a common radial factor, basically the longitudinal
form factor $F_L$. For the electric terms with $L\!=\!J$, we get
just $F_L$, while for the magnetic terms we get $\sqrt{(J+1)/(2J+1)}\;
F_L$ for $L\!=\!J\!-\!1$ and $\sqrt{J/(2J+1)}\;F_L$ for $L\!=\!J\!+\!1$.
To get the structure factors, we multiply by the OBDME with the given
(LSJ), square, and add the statistical  $(2J_f+1)/(2J_i+1)$ factor.
Note that in the case of a simple particle-hole excitation for a
closed-shell target nucleus, the $jj$ OBDME is just a phase
factor so that the (LSJ) OBDME is given by a normalized 9J
symbol for the $jj\to LS$ transformation.

 In this paper, we have just the simple $p_N\to s_\Lambda$ transition
so that for $S\!=\!0$ we need the OBDME $(101)^2$, while for 
$S\!=\!1$ we need the combination $(111)^2+3/5(112)^2$. These
are multiplied by $2\,C^2\,(2J_f+1)/(2J_i+1)$, where $C$ is the
isospin Clebsch-Gordan coefficient.


\begin{thebibliography}{99}

\bibitem{millener08} D.J.~Millener, Nucl. Phys. A 804 (2008) 84.

\bibitem{tamura08} H.~Tamura, Nucl. Phys. A 804 (2008) 73.

\bibitem{tamura10} H.~Tamura, Nucl. Phys. A 835 (2010) 3.

\bibitem{ma10} Y.~Ma et al., Nucl. Phys. A 835 (2010) 422.

\bibitem{millener10} D.J.~Millener, Nucl. Phys. A 835 (2010) 11.

\bibitem{gsd} A.~Gal, J.M.~Soper, R.H.~Dalitz, Ann. Phys. (N.Y.)
63 (1971) 53.

\bibitem{umeya11} A. Umeya, T. Harada, Phys. Rev. C 83 (2011) 034310.

\bibitem{ck65} S.~Cohen, D.~Kurath, Nucl. Phys. 73 (1965) 1.

\bibitem{millener07} D.J.~Millener, Springer Lecture Notes in Physics,
724 (2007) 31.

\bibitem{akaishi00} Y.~Akaishi, T. Harada, S. Shinmura, K.S. Myint,  
Phys. Rev. Lett. 84 (2000) 3539.

\bibitem{rijken99} Th.A.~Rijken, V.J.G.~Stoks, Y.~Yamamoto, Phys. Rev. 
C 59 (1999) 21.

\bibitem{ukai08} M.~Ukai et al., Phys. Rev. C 77 (2008) 054315.

\bibitem{ukai04} M.~Ukai et al., Phys. Rev. Lett. 93 (2004) 232501.

\bibitem{agnello09} M. Agnello et al., Phys. Lett. B 681 (2009) 139.

\bibitem{gal09} A. Gal, Nucl. Phys. A 828 (2009) 72.

\bibitem{davis63} D.H. Davis, R. Levi Setti, M. Raymund, Nucl. Phys.
41 (1963) 73; R.H. Dalitz, Nucl. Phys. 41 (1963) 78.

\bibitem{bohm74} G. Bohm et al. Nucl. Phys. B 74 (1974) 237;
D. Zieminska, R.H. Dalitz, Nucl. Phys. B 74 (1974) 248.

\bibitem{chrien90} R.E.~Chrien et al., Phys. Rev. C 41 (1990) 1062.

\bibitem{ukai06} M.~Ukai et al., Phys. Rev. C 73 (2006) 012501(R).

\bibitem{bejidian80}  M. Bejidian et al., Phys. Lett. B 94 (1980) 480.

\bibitem{mgdd85} D.J.~Millener, A.~Gal, C.B.~Dover, R.H.~Dalitz, 
Phys. Rev. C 31 (1985) 499.

\bibitem{marcucci08} L.E. Marcucci, M. Pervin, S.C. Pieper, R.
Schiavilla, R.B. Wiringa, Phys. Rev. C 78 (2008) 065501.

\bibitem{dg78} R.H.~Dalitz, A.~Gal. Ann. Phys. (N.Y.) 116 (1978) 167.

\bibitem{cusanno10} F.~Cusanno et al., Nucl. Phys. A 835 (2010) 129.

\bibitem{miwa05} K.~Miwa et al., Nucl. Phys. A 754 (2005) 80c.

\bibitem{liu88} G.-B. Liu, H.T. Fortune, Phys. Rev. C 38 (1988) 1985.

\bibitem{sotona94} M. Sotona, S. Frullani, Prog. Theor. Phys.
Suppl. 117 (1994) 151.

\bibitem{hashtam06} O. Hashimoto, H. Tamura, Prog. Part. Nucl.
Phys. 57 (2006) 564.

\bibitem{tang12} L. Tang, private communication.

\bibitem{motoba94} T. Motoba, M. Sotona, K. Itonaga, Prog. Theor. Phys.
Suppl. 117 (1994) 123.

\bibitem{davis86} D.H.~Davis, J.~Pniewski, Contemp. Phys. 27 
(1986) 91; D.H. Davis, Nucl. Phys. A 754 (2005) 3c.

\bibitem{cusanno09} F.~Cusanno et al., Phys. Rev. Lett. 103 (2009) 202501.

\bibitem{gal11} A.Gal, D.J. Millener, Phys. Lett. B 701 (2011) 342.

\bibitem{tamura12} H.~Tamura, in this issue.

\bibitem{chung76} W. Chung, PhD thesis, Michigan State University, 1976.

\bibitem{elliott99} J.P. Elliott, J.Phys. G 25 (1999) 577.

\bibitem{zofka91} J. \v{Z}ofka, L. Majling, V.N. Fetisov, R.A. Eramzhyan,
Sov. J. Part. Nucl. 22 (1991) 628.

\bibitem{majling92} L. Majling, R.A. Eramzhyan, V.N. Fetisov, 
Czech. J. Phys.  42 (1992) 1197.

\bibitem{majling97} L. Majling, R.A. Eramzhyan, V.N. Fetisov, 
Phys. Part. Nucl.  28 (1997) 101.

\bibitem{auerbach83} E.H. Auerbach et al., Ann. Phys. (N.Y.) 148
(1983) 381.

\bibitem{donnelly79} T.W. Donnelly, W.C. Haxton, At. Data Nucl. Data
Tables 23 (1979) 103.

\end{thebibliography}
\end{document}